# Quantum Engineering for Energy Applications

Florian Metzler[1], Jorge Sandoval[1], Nicola Galvanetto[1,2]

[1]Massachusetts Institute of Technology, Cambridge, Massachusetts 02139, United States.

[2]University of Zurich, Winterthurerstrasse 190, 8057 Zurich, Switzerland.

**Quantum engineering seeks to create novel technologies based on the exploitation of distinctly nonclassical behaviors such as quantum superposition. The vast majority of currently pursued applications fall into the domain of quantum information science, with quantum computing as the most visible subdomain. However, other applications of quantum engineering are fast emerging. Here, we review the deployment of quantum engineering principles in the fields of solar energy, batteries, and nuclear energy. We identify commonalities across quantum engineering approaches in those apparently disparate fields and draw direct parallels to quantum information science. We find that a shared knowledge base is forming, which de facto corresponds to a new domain that we refer to as "quantum energy science." Quantum energy science bears the promise of substantial performance improvements across energy technologies such as organic solar cells, quantum batteries, and nuclear fusion. The recognition of this emerging domain may be of great relevance to actors concerned with energy innovation. It may also benefit active researchers in this domain by increasing visibility and motivating the deployment of resources and institutional support.**

1. Introduction

Improving the performance of energy technologies is a key objective of many ongoing innovation efforts. However, prevalent energy technologies frequently run up against physical limits. An example is the ionization energy of atoms and molecules with an upper bound of around 15 eV. This number defines a hard theoretical upper limit for the achievable energy density of prevalent energy storage technologies. In actual implementations of technology, the practical limits are often much lower. Electrochemical batteries, for instance, are based on electronic redox reactions and ion migration, which constrains achievable energy density to a maximum of around 1 kWh/kg. Equivalent physical constraints apply to charging and discharging rates of batteries as well as minimum material requirements of solar cells.

Such limits can be overcome by more radically reconsidering the physics that underlies sought energy technologies. Here, we lay out how quantum engineering informs novel approaches to solar, battery and nuclear engineering – promising to overcome certain physical limitations that stifled energy technology development for decades. Such opportunities may be of interest to those in pursuit of disruptive innovation in the energy space, seeking improvements of 10x and more across key metrics. The presented approaches include high-efficiency organic solar cells with much reduced material requirements below 10 g/m$^2$ (>30x reduction compared to present levels), quantum batteries with enhanced charging through superabsorption (>100x acceleration), and high-capacity nuclear batteries that use atomic nuclei as energy reservoirs (>1000x increase). See Supplementary Note S1 for supporting calculations.

While implementations may look different across domains, there is a common denominator to quantum-engineering-based energy technologies: in each case, energy is deliberately moved into and out of energetic states of quantum systems, which can comprise molecules, atoms, and nuclei. Control over this process is achieved by selectively activating or suppressing couplings between such quantum systems. As a result of such deliberate control, new fundamental building blocks become available to be utilized in energy technologies; and collective quantum effects can be employed to greatly accelerate desirable dynamics. The employed theoretical and practical toolsets are closely related to those of other quantum engineering domains, providing ample opportunities for spillover of knowledge and human resources.

Quantum computing as well as quantum simulation, quantum sensing, and quantum communication are the most visible applications of quantum engineering today. Around the globe, large programs are forming to systematically support and foster an emerging quantum ecosystem (ref.). However, energy applications are typically missing from such programs. We argue that the vast potential of quantum engineering for energy applications warrants greater attention to that domain and suggest its inclusion in major quantum initiatives.

In the remainder of this article, we provide an overview of quantum engineering for energy applications. We introduce key principles, lay out the possibilities that researchers have envisioned, and summarize what has been accomplished to date.

## 2. Background and key principles

The properties and behaviors of most of our materials are governed by quantum principles. These insights, first recognized during the early 20$^{th}$ century, led to the first quantum revolution, which revolved around aggregate effects of quantum principles such as the electronic band structure of materials. During the 1970s, a second quantum revolution emerged and is today in full swing. The new wave of quantum engineering associated with it – sometimes referred to as Quantum 2.0 – focuses on the precise control of quantum properties and on the exploitation of distinctly nonclassical behaviors such as superposition and entanglement (Pritchard and Till 2014; Awschalom et al. 2017; Rapp & Schneider 2021). Quantum computing is perhaps the most visible manifestation of Quantum 2.0. In quantum computing, quantum systems with discrete energetic states such as atoms and small superconducting circuits are deliberately excited and then allowed to time-evolve in probabilistically predictable ways.

The storage and manipulation of information entails the storage and manipulation of energy – a principle that applies to classical physics and quantum physics alike (Szilard 1929). Consider a memory chip where the information content of a binary bit is represented through a small amount of charge. For information applications, engineers seek to minimize the amount of charge required while maximizing the fidelity with which it can be retrieved**.** Priorities differ for energy applications: Consider a bank of capacitors which can be either empty (0) or charged (1). Here, engineers care more about the capacity of the system than about fidelity. The same shift of perspective applies to quantum science. Two-level quantum systems, widely known as *quantum bits* or *qubits*, can be in a higher energetic state or in a lower energetic state – which can be interpreted as |1> and |0> when viewed from an information perspective, or as "charged" and "uncharged" when viewed from an energy perspective (Hübler and Osuagwu 2010). Quantum systems that are optimized for energy storage, transfer and conversion rather than for information storage, transfer and conversion have been referred to as *work qubits* (Binder et al. 2015).

What both information and energy perspectives have in common is that quantum systems can be manipulated by temporarily enhancing interactions between them (Majer et al. 2007). This allows for energy to redistribute and energetic states to change. As a result of enhanced interactions, quantum systems that are now coupled can no longer be seen as separate entities. Like multiple small waves forming a single larger one, energy is then – for a short period of time – held collectively in an overarching system, where it can no longer be pinpointed. This collective state is known as a state of coherent superposition (Vewinger et al. 2003; Sillanpää et al. 2007). Eventually, the superposition state ends – often within nanoseconds – due to disturbances from the environment. Energy may then end up redistributed compared to the system's initial state. This redistribution effectively represents an energy transfer or an energy conversion event. The same process can also be interpreted as an information transfer or information conversion event – as is the case in quantum state transfer between qubits (Fig. 1a). Fig. 1b shows two coupled qubits and Fig. 1c shows six coupled qubits. In both cases, the interactions between the qubits can be tuned via external stimulation, so as to accelerate or decelerate their excitation and deexcitation through induced energy transfers via superposition states (Wu et al. 2018; Bernien et al. 2017).

Several characteristics are important here. First, this quantum mode of energy redistribution – also known as quantum energy transfer or quantum transport – can be extremely fast (Jones & Bradshaw 2019). Consequently, quantum energy transfer allows for the preclusion of competing events – such as the spontaneous decay of an excited molecule. Fast dynamics are in part driven by collective quantum effects known as Dicke enhancement (also known as superradiance, superabsorption, and supertransfer – see Dicke 1956; Lloyd & Mohseni 2010; Higgins et al. 2014). Second, the rate at which energy transfer occurs depends on the extent to which resonances exist in the coupled system (Andrews 2009). Providing resonant acceptors – matched by design or via in situ tuning – then allows for the acceleration of desirable transitions. Finally, this mode of energy redistribution is in principle lossless – in contrast to other modes of energy transfer that dissipate energy such as moving electrons.

To better understand these unique features of manipulating energy at the nanoscale, researchers have created models that describe relevant aspects of quantum systems and their interactions. Such models can then be evolved over time to study their dynamic behavior. A common way of representing these models is through Hamiltonians – equations that keep track of a system's energy reservoirs and energy transfer pathways, along with the factors that affect them (Eisberg & Resnick 1985). As laid out in Fig. 1 and further described in Supplementary Note S2, the models – and the quantum principles that govern them – are highly similar across the different cases discussed in this article: solar energy, batteries, and nuclear energy.

In the following sections, we will provide examples of emerging quantum engineering applications in each domain and discuss their potential for the improvement of existing or the creation of novel energy technologies.

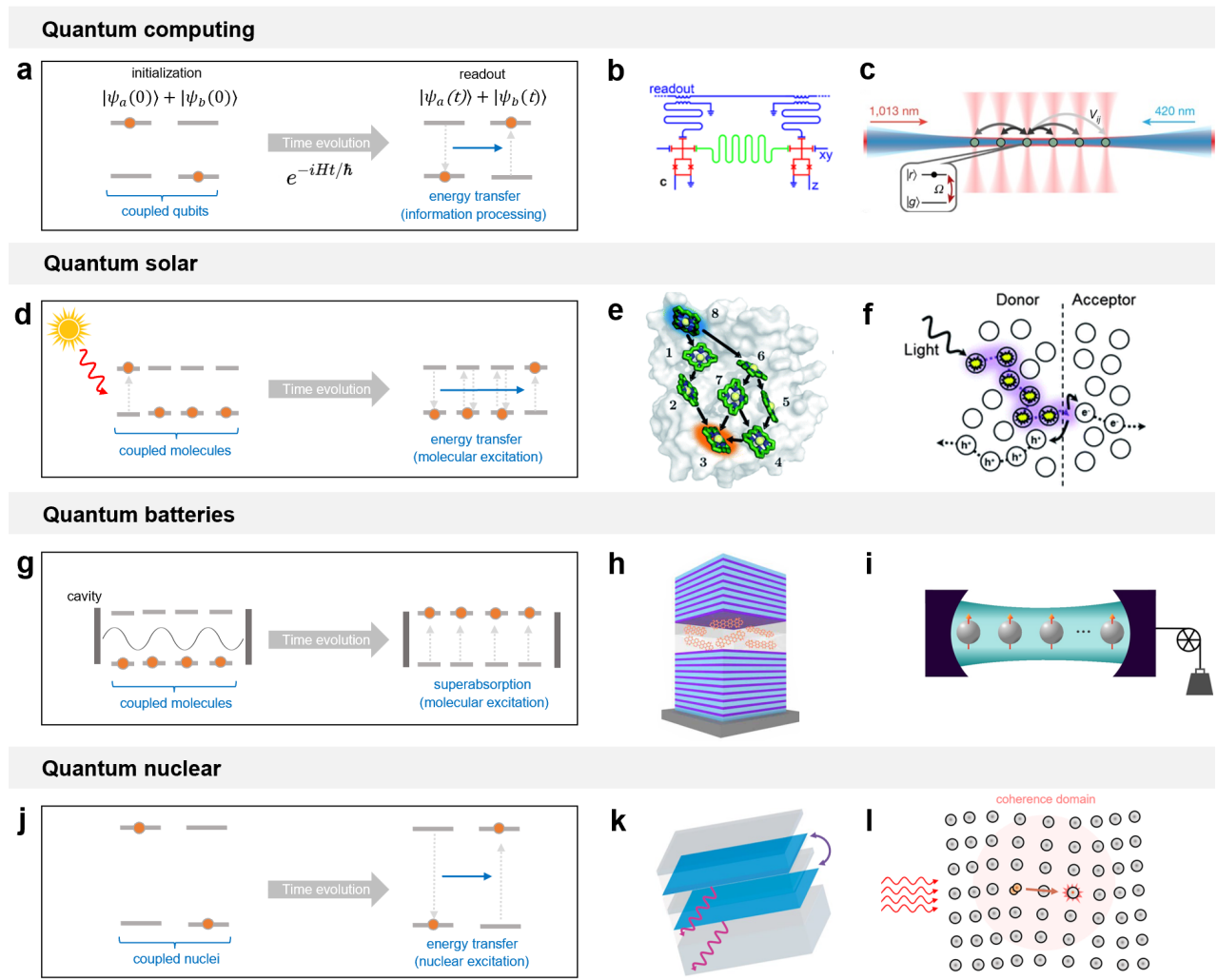

Fig. 1: **a** Schematic depiction of the time evolution of two qubits that results in quantum state transfer. Superposition states are initiated by increasing couplings between qubits via external stimulation. **b** Implementation of coupled qubits via superconducting circuits (red) and a shared oscillator (green), as shown in Wu et al. 2018. **c** Implementation of coupled qubits via Rydberg atoms held by optical tweezers (vertical beams) where interactions can be tuned via shared photonic modes (horizontal beams), as shown in Bernien et al. 2017. **d** Schematic depiction of molecular exciton transfer. Superposition states are initiated by dipole-dipole coupling between molecules. **e** Implementation of molecular exciton transfer in a photosynthetic system (Fenna–Matthews–Olson complex), as shown in Häse et al. 2020. **f** Molecular exciton transfer in an organic solar cell where excitation transfers from a donor system in the absorption layer to an acceptor system at the charge separation barrier, as shown in Menke & Holmes 2014. **g** Schematic depiction of charging, via superabsorption, a quantum battery, which consists of multiple work qubits. Superposition states are initiated by increasing couplings between the work qubits via external stimulation. **h** Implementation of a quantum battery via organic molecules (Lumogen-F orange) driven by a cavity, as shown in Quach et al. 2022. **i** Illustration of a quantum battery as an array of work qubits driven by a cavity, as shown in Binder et al. 2015 **j** Schematic depiction of the time evolution of two nuclear work qubits that results in quantum state transfer. Superposition states are initiated by increasing couplings between the work qubits via external stimulation. **k** Implementation of coupled nuclear work qubits via thin-film $^{57}Fe$ layers that act as resonant cavities, as shown in Haber et al. 2017. **l** Proposed implementation of coupled nuclear work qubits with deuteron pairs as donor systems and high-Z lattice nuclei as resonant acceptor systems, as described in Hagelstein & Chaudhary 2015.

## 3. Subdomains of quantum energy science
### 3.1. Quantum solar

Quantum engineering is now employed to apply lessons learned from photosynthesis to solar cell design to maximize the efficiency of light harvesting. The aspiration is to combine nature's high efficiency in utilizing absorbed photons with the ability of artificial materials to absorb a greater number of photons (by making use of a wider range of the light spectrum compared to biological systems; Brédas et al. 2017).

The extremely high efficiency in the utilization of absorbed photons in photosynthesis has startled scientists for decades (Cao et al. 2020). Almost every photon that is absorbed by a photosynthetic system is used in the formation of chemical bonds, rather than getting reemitted and lost. Achieving photon utilization rates close to 100% is not trivial since the excited molecular-level quantum systems that result from photon absorption – known as excitons – are extremely short-lived (Dursun & Guzelturk 2021). If the energy held by an exciton is not moved within picoseconds to a reaction center molecule, where sugar production occurs, then the exciton decays and the excitation energy is reemitted in the form of a photon. If plants were to employ classical transfer mechanisms e.g. where energy to be transferred is imparted onto a particle that physically moves through its environment akin to the migration of ions in batteries, then most of the initially absorbed energy would be lost due to the slowness of such a process. In actuality, it is not energetic particles that move but only the excitation energy itself, in an oscillator-mediated, nonradiative transfer process via the temporary superposition and delocalization of coupled molecules (Andrews et al. 2020; see Fig. 1d and 1e). Finally, because of boundary conditions in the photosynthetic complex, the energy in the system is more likely to end up – when the coherent superposition state is broken – at the reaction center of the complex, where it needs to be. This efficient path-finding process – known as a quantum walk – has been compared to plants engaging in a form of quantum computing (Biello 2007; Mohseni et al. 2008). When viewed from this perspective, excitons can be seen as excited work qubits that temporarily hold energy – energy which then transfers to other work qubits as the exciton diffuses.

Theoretical models of this energy transfer process suggest that it works so well because parameters of the surrounding environment are just right, ensuring maximum efficiency. Apparently nature has evolved ways of maintaining temporary superposition states despite the presence of noisy environments – and noisy environments may even contribute to the stabilization of such states (Plenio & Huelga 2008). Insights from such models can then be applied to the design of technology (Mattioni et al. 2021). Specifically, solar cells face similar challenges as photosynthetic systems: if energy is not transferred fast enough from the sites of exciton creation (the top layer of a solar cell where photons are absorbed) to the sites where excitons generate charge carriers (the bottom layer of a solar cell), then energy is lost through reemission (Menke & Holmes 2014; also see Fig. 1f). In practice, this means that the top layer of the cell is limited in thickness to the exciton diffusion length i.e. the distance that excitons manage to travel within their short lifetime. However, thin top layers, e.g. with thicknesses on the order of a few nm, mean that much of the incoming light is not utilized, as many photons require more than 100 nm of penetration depth to be absorbed (Sajjad et al. 2020).

This is of particular concern with organic solar cells, one of the most promising types of emerging solar technologies. Organic solar cells – sometimes called plastic solar cells – are made from polymeric materials, making them low-cost in their fabrication and installation (Chiechi et al. 2013). Other advantages are low material requirements (<10 g/sqm), environmentally friendly disposal, bendability, and the possibility to create semi-transparent cells – all factors that enable diverse and broad deployment. The key drawback to date has been the comparatively low efficiency of organic solar cells – often below 10% – especially when paired with practical considerations such as large-scale production and longevity. The low efficiencies are rooted in the short lifetimes of excitons in organic materials, which correspond to short diffusion lengths (Riede et al. 2021). When working with thick absorption layers >10 nm, many excitons recombine and lose their energy to the environment before they can reach a junction where this excitation energy is converted to charge. In certain prototype cells, this issue has been addressed through the creation of special structures where junctions with charge separation barriers permeate the absorption layer, known as *bulk heterojunctions* (Rand and Richter 2014). However, such designs are not as scalable and robust as the much simpler *planar heterojunctions*, which are essentially plane layers of thin film. To achieve both high efficiency and high manufacturability as well as robustness, researchers aim for planar heterojunctions with long exciton diffusion lengths across deep absorption layers. Ideally, organic solar cells exhibit absorption layers *and* corresponding diffusion lengths in excess of 100 nm.

Increases of exciton diffusion lengths in organic materials have been achieved by testing different material configurations, without fully understanding why some work better than others (Cnops et al. 2014; Long et al. 2016; Sneyd et al. 2021). Sneyd et al. 2021, for instance, report in films of oligomeric polyfluorene nanofibers an exciton diffusion length of 300 nm +/- 50 nm, estimated from pump-probe measurements. To truly optimize exciton diffusion in a variety of materials, however, researchers seek to identify ideal material configurations and morphologies based on models and simulations of the above-described nonradiative transfer processes, which would then allow for the prediction of optimal parameters (Ruseckas & Samuel 2021; Mattioni et al. 2021).

To summarize: quantum models have been developed to explain the remarkably high efficiency of energy transfer in photosynthesis. These models suggest the presence of the same coupling-induced superposition states that are used in quantum computing. The analogy becomes particularly clear when viewing excited molecules as work qubits. The interactions between such work qubits allow for fast excitation transfer across them. In quantum solar engineering, the described models inform optimal parameters for the design of new solar cell materials that seek similarly efficient energy transfer performance. Plants have an internal efficiency of close to 100% – meaning that almost all absorbed photons are utilized. However, because only small parts of the light spectrum are utilized, their overall efficiency is only about 1% – meaning most of the available photons are not absorbed. Combining the best of both worlds, namely absorbing a much larger portion of the light spectrum while engineering a high internal efficiency through fast excition diffusion, is sought to lead to robust, scalable, and low-cost organic solar cells with overall efficiencies in excess of 20%. 20% is the estimated threshold for large-scale adoption of organic solar cells, as estimated by Riede et al. 2021. A discussion of an exemplary Hamiltonian that has been used by researchers to model Dicke-enhanced exciton diffusion (Lloyd & Mohseni 2010; Abasto et al. 2012) can be found in Supplementary Note S3.

### 3.2. Quantum batteries

The equivalence of energy and information led researchers to consider whether two-level quantum systems, i.e. qubits, could be optimized for the storage and retrieval of energy instead of the storage and retrieval of information. In this case, design requirements shift: addressability of individual qubits is not required; however, the energy held per qubit ought to be high and the total number of qubits large.

In one proposed design for quantum batteries, energy is stored in excited states of dye molecules (Lumogen-F Orange) which are contained in a thin film polymer matrix (Fig. 1h; Quach et al. 2022). Data from one such experimental configuration suggest that $1.6 \times 10^{11}$ molecules were successfully used as work qubits, each with an energy transition of 0.108 eV (resulting in a total capacity of the battery prototype of $2.768 \times 10^{-9}$ J). Charging occurs through exciting work qubits via coherent laser light, although the authors of the study suggest that in the future, incoherent sunlight might be used as well. An alternative configuration is proposed by Cruz et al. 2021, which uses as work qubits molecules known as metal carboxylates. These authors report the storage of 0.85 mJ of energy per mol of the molecule $Cu_2(HCOO)_4(HCOOH)_2(C_4H_{10}N_2)$. The above values are still far below the energy densities of common batteries (which are in the hundreds of Wh/kg range, whereas the values of Cruz et al. correspond to about 1 µWh/kg and the values of Quach et al. correspond to about 4 Wh/kg). However, these early experimental configurations already exhibit some key strengths of anticipated quantum batteries. In Quach et al., remarkable power densities – i.e. charging and discharging rates – such as 67 kW/kg are reported from experiments. This value already far exceeds the power densities of existing lithium-ion batteries, which remain below 900 W/kg (see Supplementary Note S1). Even the most advanced supercapacitors do not exceed a power density of 10 kW/kg.

While such early experimental reports need to be treated with caution, they already exhibit a key principle motivating quantum battery design: when the transfer of energy to and from the work qubits occurs through superposition states, then the process can be very fast. Specifically, the dynamics benefit from quantum speedup associated with Dicke enhancement (superradiance and its inverse mechanism, superabsorption). This means that, depending on the exact configuration, charging and discharging times can scale by a factor of $\sqrt{N}$, N, or $N^2$, where N is the number of work quits that participate in the intermediate superposition state (Higgins et al. 2014; Campaioli et al. 2018; Ferraro et al. 2018; Gumberidze et al. 2019). In other words, the larger the battery (and the coherence state that can form within it), the faster it can charge and discharge. Quantum battery designers then seek to optimize such configurations toward achieving both high power densities and high energy densities. Higher energy densities, compared to those demonstrated to date, are conceivable when quantum systems with larger transition energies and smaller molecular weights are utilized.

Like in quantum solar engineering, system designs are informed by Hamiltonian-based models that describe the major components and couplings in such systems (see Supplementary Note S4). Evaluating predicted system performance across the design space then informs the optimal composition and structure of the employed materials. As a result, key metrics such as charging rates, energy retention, and discharging rates can be optimized.

When tuned sufficiently well, charging and discharging rates for such batteries can be almost instantaneous (Alicki & Fannes 2013). Resulting technology could overcome a major limitation of existing battery technology where charging and discharging is severely rate-limited, a constraint that has to date impeded battery deployment in such mass-market applications as vehicles and airplanes as well as such specialized applications as high-end lasers and pulsed magnetic fields.

### 3.3. Quantum nuclear

Thus far, we have considered as relevant two-level systems – i.e. as work qubits – quantum systems at the atomic level such as molecules and atoms. Atomic nuclei, too, can absorb and emit discrete quanta of energy and therefore also fulfill the basic condition for qubits or work qubits (Boyles 2014). This prompted some researchers to consider the use of nuclei as work qubits for quantum batteries – what could then be called nuclear quantum batteries (Liao 2014).

The prospective advantage of nuclear quantum batteries is vastly increased energy density, exceeding by orders of magnitude any current storage technology. This is because nuclear excited states can typically accommodate more than 100,000 times the energy of atomic and molecular excited states (Supplementary Note S1). However, interactions that at the atomic level can readily cause superposition states and alternative energy transfer pathways are much weaker at the nuclear level. Nevertheless, researchers have succeeded experimentally in creating nuclear superposition states, and in increasing and decreasing the decay rates of such nuclear excited states (Chumakov et al. 2018; Heeg et al. 2021) – important steps toward the coherent control of nuclei. In Chumakov et al. 2018, for instance, the acceleration of the 14 keV $|^{57}Fe^*\rangle \rightarrow |^{57}Fe\rangle$ nuclear de-excitation by a factor of 15 is shown experimentally. Like in photosynthesis and in molecular quantum batteries, at the nuclear scale too, conventionally slow dynamics can be accelerated through Dicke enhancement (Terhune & Baldwin 1965; Hannon & Trammell 1999). Accelerated excitation and deexcitation correspond to charging and discharging processes in quantum batteries; and delayed deexcitation corresponds to retaining stored energy.

Other research shows that relevant interactions between oscillators and nuclei and between nuclei and nuclei can be increased by external stimulation such as externally applied fields (Haber et al. 2017; Bocklage et al. 2021). Fig. 1j shows an experimental configuration where nuclear excitation energy was transferred through external laser stimulation between two collective nuclear excited states (also known as nuclear excitons – note the similarity in nomenclature to molecular excitons used in solar engineering). Varying the parameters of such external stimulation can moreover be used to tune systems into and out of resonance, further accelerating or decelerating sought dynamics. Such considerations and early results motivate the nascent field of quantum nuclear engineering.

The potential scope of quantum nuclear engineering becomes even larger when considering its possible effect on nuclear reactions. Some researchers proposed that principles applied to the coherent control of nuclei may not only apply to nuclear state transitions but also to nuclear reactions[1] (Schwinger 1990; Preparata 1995; Hagelstein & Chaudhary 2015). In this picture, if nuclear reactions take place in an environment where a temporary superposition of multiple nuclei is achieved – e.g. due to external stimulation – then the resulting energy redistribution within the coupled system affects the nuclear reaction parameters. For instance, if the energy release associated with one nuclear reaction can be readily absorbed by a counterpart reaction in the same coupled system – in other words, if the two complementary reactions are resonant – then the rate at which those reactions occur is expected to increase, compared to the isolated case (Hagelstein et al. 2009; Hagelstein & Chaudhary 2015).

Specifically, Schwinger (1994) proposed to view a pair of deuterium nuclei $D_2$ as a four-nucleon quantum system in an excited state, i.e. as the equivalent of a work qubit in state $|1\rangle$. Transitioning to a more compact and less energetic four-nucleon configuration such as a stable $^4He$ nucleus then corresponds to a deexcitation step – analogous to the induced flip of a qubit from $|1\rangle$ to $|0\rangle$ (Fig. 1a). Researchers have proposed that this dynamic could be accelerated by stimulating interactions between deuteron pairs and matching heavy nuclei that exhibit excited states resonant with the 23.8 MeV $|D_2\rangle \rightarrow |^4He\rangle$ deexcitation process (Fig. 1k). Such a mechanism would be analogous to the accelerated exciton diffusion (Fig. 1f) and analogous to the accelerated charging and discharging of a quantum battery's work qubits (Fig. 1h). It would represent the acceleration of nuclear reactions beyond their conventionally expected reaction rates. Researchers envision using such techniques to accelerate the decay of nuclear waste and to render desirable fusion reactions more accessible (Dumé 2021; Metzler et al. 2022).

As in the examples above, the proposed systems and processes – from nuclear quantum batteries to accelerated nuclear reactions – can be modeled via Hamiltonian-based models that capture relevant features of the participating quantum systems and the interactions between them (see Supplementary Notes S5 and S6). The time evolution of such models then identifies dominant energy transfer pathways and rates. The dependence of such dynamics on environmental parameters such as geometric arrangements and on stimulation characteristics informs the design of materials and promises the rational engineering of corresponding technologies.

---

[1] From this perspective, nuclear reactions are viewed as a subset of nuclear state transitions. Nuclear state transitions in general terms are reconfigurations of nucleons hindered by barriers that are overcome probabilistically (akin to how atomic state transitions are reconfigurations of electrons). Nuclei relaxing from excited states to lower energy states then correspond to the reconfiguration of nucleons within nuclei, whereas nuclear reactions such as fusion reactions correspond to the reconfiguration of nucleons across nuclei.

*Table 1: Overview of presented quantum engineering domains*

| Domain | Choice of work qubits | Experimental results to date | Promise |
|---|---|---|---|
| Quantum solar | Molecules and molecular excitons (electron-hole pairs) | Molecular exciton diffusion lengths of >300 nm demonstrated; acceleration of deexcitation (through supertransfer) by a factor >100x demonstrated. See Sneyd et al. 2021; Mattioni et al. 2021. | Low-cost and robust organic solar cells with high efficiency; utilization of a broader range of the light spectrum |
| Quantum batteries | Molecules and atoms | Acceleration of excitation and deexcitation of molecules (through superabsorption and superradiance) by a factor of >1000x demonstrated. Charging rates up to 854 kW/kg reported (at a charging capacity of about 4 Wh/kg). See Quach et al. 2021; Cruz et al. 2021. | Batteries with quasi-instantaneous charging and discharging; lossless storage through 'dark states' |
| Quantum nuclear | Nuclei and nuclear excitons (collective nuclear excitations) | Deliberately induced deexcitation of nuclear states in $^{57}$Fe; acceleration of nuclear deexcitation (through superradiance) by 15x demonstrated; phase alignment of nuclear states across multiple nuclei via external radio frequency stimulation demonstrated. See Haber et al. 2017; Chumakov et al. 2018; Bocklage et al. 2021. | Nuclear quantum batteries with high storage capacity; nuclear waste processing through accelerated nuclear decay, compact fusion technology through accelerated nuclear reactions |

## 4. Conclusions

We described the application of quantum science and quantum engineering principles to energy technologies, across the domains of solar technology, battery technology, and nuclear technology. An overview is provided in Table 1. In each of these domains, the quantum-based approaches described here are to date largely viewed as niche areas. However, the knowledge bases that these approaches draw on tend to have more in common with each other than with the traditional knowledge bases of each of their corresponding application domains.

We have seen that there is a common denominator across the presented cases, namely modeling different kinds of quantum systems through appropriate Hamiltonians and then studying the time evolution and energy redistribution as a function of key parameters of the respective physical configuration. This then informs the design of new systems – or the deliberate modification of existing systems – and enables the optimization of relevant parameters.

Different names have been proposed to describe this kind of modeling and engineering process: "energy transfer editing" (Qin et al. 2019), "coherence engineering" (Huynh et al. 2012), and "transition rate engineering" (Higgins et al. 2014) are just a few examples. Broader terms include "tuning chemical reactions" (Tan et al. 2017) and "tuning nuclear reactions" (Metzler et al. 2022).

We posit that the shared knowledge base across the different energy applications of quantum engineering justifies the recognition as a dedicated subfield of quantum science, which may be referred to as "quantum energy science" as a counterpart to "quantum information science".

Research on traditional energy technologies and research on quantum information science both enjoy high public interest and high levels of investment. Recognizing quantum energy science as a dedicated field that emerges at the intersection of these two major domains could lead to greater awareness of its potential societal impact, and greater support in the form of human and financial resources.

**Acknowledgments:**

Support from the Anthropocene Institute is gratefully acknowledged.

**References:**

Alicki, R., & Fannes, M. (2013). Entanglement boost for extractable work from ensembles of quantum batteries. *Physical Review E*, *87*(4), 042123.

Andrews, D. L. (2009). *Resonance energy transfer: Theoretical foundations and developing applications*. SPIE Press: Washington.

Andrews, D. L., Bradshaw, D. S., Forbes, K. A., & Salam, A. (2020). Quantum electrodynamics in modern optics and photonics. *J. Opt. Soc. Am*.

Awschalom, D., Christen, H., Clerk, A., Denes, P., Flatté, M., Freedman, D., Galli, G., Jesse, S., Kasevich, M., & Monroe, C. (2017). *Basic Energy Sciences Roundtable: Opportunities for Basic Research for Next-Generation Quantum Systems*. USDOE Office of Science (SC)(United States).

Bernien, H., Schwartz, S., Keesling, A., Levine, H., Omran, A., Pichler, H., Choi, S., Zibrov, A. S., Endres, M., Greiner, M., Vuletić, V., & Lukin, M. D. (2017). Probing many-body dynamics on a 51-atom quantum simulator. *Nature*, *551*(7682), 579–584.

Biello, D. (2007). When it comes to photosynthesis, plants perform quantum computation. *Scientific American April*, *13*.

Binder, F. C., Vinjanampathy, S., Modi, K., & Goold, J. (2015). Quantacell: Powerful charging of quantum batteries. *New Journal of Physics*, *17*(7), 075015.

Bocklage, L., Gollwitzer, J., Strohm, C., Adolff, C. F., Schlage, K., Sergeev, I., Leupold, O., Wille, H.-C., Meier, G., & Röhlsberger, R. (2021). Coherent control of collective nuclear quantum states via transient magnons. *Science Advances*, *7*(5), eabc3991.

Boyles, A. (2014, March 20). Gamma-ray shaping could lead to "nuclear" quantum computers. Physics World. ttps://physicsworld.com/a/gamma-ray-shaping-could-lead-to-nuclear-quantum-computers/

Brédas, J.-L., Sargent, E. H., & Scholes, G. D. (2017). Photovoltaic concepts inspired by coherence effects in photosynthetic systems. *Nature Materials*, *16*(1), 35–44.

Campaioli, F., Pollock, F. A., & Vinjanampathy, S. (2018). Quantum batteries. In *Thermodynamics in the Quantum Regime* (pp. 207–225). Springer.

Cao, J., Cogdell, R. J., Coker, D. F., Duan, H.-G., Hauer, J., Kleinekathöfer, U., Jansen, T. L. C., Mančal, T., Miller, R. J. D., Ogilvie, J. P., Prokhorenko, V. I., Renger, T., Tan, H.-S., Tempelaar, R., Thorwart, M., Thyrhaug, E., Westenhoff, S., & Zigmantas, D. (2020). Quantum biology revisited. *Science Advances*, *6*(14), eaaz4888.

Chiechi, R. C., Havenith, R. W., Hummelen, J. C., Koster, L. J. A., & Loi, M. A. (2013). Modern plastic solar cells: Materials, mechanisms and modeling. *Materials Today*, *16*(7–8), 281–289.

Chumakov, A. I., Baron, A. Q. R., Sergueev, I., Strohm, C., Leupold, O., Shvyd'ko, Y., Smirnov, G. V., Rüffer, R., Inubushi, Y., Yabashi, M., Tono, K., Kudo, T., & Ishikawa, T. (2018). Superradiance of an ensemble of nuclei excited by a free electron laser. *Nature Physics*, *14*(3), 261–264.

Cnops, K., Rand, B. P., Cheyns, D., Verreet, B., Empl, M. A., & Heremans, P. (2014). 8.4% efficient fullerene-free organic solar cells exploiting long-range exciton energy transfer. *Nature Communications*, *5*(1), 3406.

Cruz, C., Anka, M. F., Reis, M. S., Bachelard, R., & Santos, A. C. (2021). Quantum battery based on quantum discord at room temperature. *ArXiv:2104.00083 [Quant-Ph]*. http://arxiv.org/abs/2104.00083

Dicke, R. H. (1954). Theory of superradiance. *Physical Review*, *93*(1), 99–110.

Dumé, I. (2021, March 4). *Atomic nuclei go for a quantum swing*. Physics World. https://physicsworld.com/atomic-nuclei-go-for-a-quantum-swing/

Dursun, I., & Guzelturk, B. (2021). Exciton diffusion exceeding 1 μm: Run, exciton, run! *Light: Science & Applications*, *10*(1), 39.

Eisberg, R., & Resnick, R. (1985). *Quantum physics of atoms, molecules, solids, nuclei, and particles*.

Ferraro, D., Campisi, M., Andolina, G. M., Pellegrini, V., & Polini, M. (2018). High-Power Collective Charging of a Solid-State Quantum Battery. *Physical Review Letters*, *120*(11), 117702.

Gumberidze, M., Kolář, M., & Filip, R. (2019). Measurement Induced Synthesis of Coherent Quantum Batteries. *Scientific Reports*, *9*(1), 1–12.

Haber, J., Kong, X., Strohm, C., Willing, S., Gollwitzer, J., Bocklage, L., Rüffer, R., Pálffy, A., & Röhlsberger, R. (2017). Rabi oscillations of X-ray radiation between two nuclear ensembles. *Nature Photonics*, *11*(11), 720–725.

Hagelstein, P. L., & Chaudhary, I. U. (2015). Phonon models for anomalies in condensed matter nuclear science. *Current Science*, 507–513.

Hagelstein, P. L., Chaudhary, I. U., McKubre, M. C. H., & Tanzella, F. (2009). Progress toward a theory for excess heat in metal deuterides. *AIP Conference Proceedings*, *1154*(1), 257–271.

Hannon, J. P., & Trammell, G. T. (1999). Coherent γ-ray optics. *Hyperfine Interactions*, *123*(1), 127–274.

Häse, F., Roch, L. M., Friederich, P., & Aspuru-Guzik, A. (2020). Designing and understanding light-harvesting devices with machine learning. *Nature Communications*, *11*, 4587.

Heeg, K. P., Kaldun, A., Strohm, C., Ott, C., Subramanian, R., Lentrodt, D., Haber, J., Wille, H.-C., Goerttler, S., Rüffer, R., Keitel, C. H., Röhlsberger, R., Pfeifer, T., & Evers, J. (2021). Coherent X-ray−optical control of nuclear excitons. *Nature*, *590*(7846), 401–404.

Higgins, K. D. B., Benjamin, S. C., Stace, T. M., Milburn, G. J., Lovett, B. W., & Gauger, E. M. (2014). Superabsorption of light via quantum engineering. *Nature Communications*, *5*(1), 4705.

Hübler, A. W., & Osuagwu, O. (2010). Digital quantum batteries: Energy and information storage in nanovacuum tube arrays. *Complexity*, *15*(5), 48–55.

Huynh, P.-A., Portier, F., le Sueur, H., Faini, G., Gennser, U., Mailly, D., Pierre, F., Wegscheider, W., & Roche, P. (2012). Quantum Coherence Engineering in the Integer Quantum Hall Regime. Physical Review Letters, 108(25), 256802.

Jones, G., & Bradshaw, D. (2019). Resonance energy transfer: From fundamental theory to recent applications. *Frontiers in Physics*, *7*.

Liao, W.-T. (2014). *Coherent Control of Nuclei and X-Rays*. Springer Science & Business Media.
12

Lloyd, S., & Mohseni, M. (2010). Symmetry-enhanced supertransfer of delocalized quantum states. *New Journal of Physics*, *12*(7), 075020.

Long, G., Wu, B., Solanki, A., Yang, X., Kan, B., Liu, X., Wu, D., Xu, Z., Wu, W.-R., & Jeng, U.-S. (2016). New insights into the correlation between morphology, excited state dynamics, and device performance of small molecule organic solar cells. *Advanced Energy Materials*, *6*(22), 1600961.

Majer, J., Chow, J. M., Gambetta, J. M., Koch, J., Johnson, B. R., Schreier, J. A., Frunzio, L., Schuster, D. I., Houck, A. A., Wallraff, A., Blais, A., Devoret, M. H., Girvin, S. M., & Schoelkopf, R. J. (2007). Coupling superconducting qubits via a cavity bus. *Nature*, *449*(7161), 443–447.

Mattioni, A., Caycedo-Soler, F., Huelga, S. F., & Plenio, M. B. (2021). Design Principles for Long-Range Energy Transfer at Room Temperature. *Physical Review X*, *11*(4), 041003.

Menke, S. M., & Holmes, R. J. (2014). Exciton diffusion in organic photovoltaic cells. *Energy & Environmental Science*, *7*(2), 499–512.

Metzler, F., Hunt, C., & Galvanetto, N. (2022). Known mechanisms that increase nuclear fusion rates in the solid-state (arXiv:2208.07245). *arXiv*. https://doi.org/10.48550/arXiv.2208.07245

Mohseni, M., Rebentrost, P., Lloyd, S., & Aspuru-Guzik, A. (2008). Environment-assisted quantum walks in photosynthetic energy transfer. *The Journal of Chemical Physics*, *129*(17), 174106.

Plenio, M. B., & Huelga, S. F. (2008). Dephasing-assisted transport: Quantum networks and biomolecules. *New Journal of Physics*, *10*(11), 113019.

Preparata, G. (1995). QED Coherence in matter. World Scientific.

Pritchard, J., & Till, S. (2014). *UK Quantum Technology Landscape 2014* (Defence Science and Technology Laboratory, DSTL/PUB75620).

Quach, J. Q., McGhee, K. E., Ganzer, L., Rouse, D. M., Lovett, B. W., Gauger, E. M., Keeling, J., Cerullo, G., Lidzey, D. G., & Virgili, T. (2022). Superabsorption in an organic microcavity: Toward a quantum battery. *Science Advances*, *8*(2), eabk3160.

Rand, B. P., & Richter, H. (2014). *Organic Solar Cells: Fundamentals, Devices, and Upscaling*. CRC Press.

Rapp, H. P., & Schneider, S. (2021). *Economic-technological revolution through Quantum 2.0* [Deutsche Bank Research Report].

Riede, M., Spoltore, D., & Leo, K. (2021). Organic Solar Cells—The Path to Commercial Success. *Advanced Energy Materials*, *11*(1), 2002653.

Ruseckas, A., & Samuel, I. D. (2021). Engineering highways for excitons. *Joule*, *5*(11), 2762–2764.

Sajjad, M. T., Ruseckas, A., & Samuel, I. D. (2020). Enhancing Exciton Diffusion Length Provides New Opportunities for Organic Photovoltaics. *Matter*, *3*(2), 341–354.

Schwinger, J. (1990). Nuclear energy in an atomic lattice. *Zeitschrift Für Physik D Atoms, Molecules and Clusters*, *15*(3), 221–225.

Schwinger, J. (1994). Cold Fusion Theory–A Brief History of Mine. *Fusion Technology*, *26*(4), XIII.

Sillanpää, M. A., Park, J. I., & Simmonds, R. W. (2007). Coherent quantum state storage and transfer between two phase qubits via a resonant cavity. *Nature*, *449*(7161), 438–442.

Sneyd, A. J., Fukui, T., Palecek, D., Prodhan, S., Wagner, I., Zhang, Y., Sung, J., Collins, S. M., Slater, T. J. A., Andaji-Garmaroudi, Z., MacFarlane, L. R., Garcia-Hernandez, J. D., Wang, L., Whittell, G. R.,

Hodgkiss, J. M., Chen, K., Beljonne, D., Manners, I., Friend, R. H., & Rao, A. (n.d.). Efficient energy transport in an organic semiconductor mediated by transient exciton delocalization. *Science Advances*, 7(32), eabh4232.

Szilard, L. (1929). Über die Entropieverminderung in einem thermodynamischen System bei Eingriffen intelligenter Wesen. *Zeitschrift Für Physik*, *53*(11), 840–856.

Tan, S., Xia, T., Shi, Y., Pfaendtner, J., Zhao, S., & He, Y. (2017). Enhancing the Oxidation of Toluene with External Electric Fields: A Reactive Molecular Dynamics Study. Scientific Reports, 7(1), 1710.

Terhune, J. H., & Baldwin, G. C. (1965). Nuclear Superradiance in Solids. *Physical Review Letters*, *14*(15), 589–591.

Vewinger, F., Heinz, M., Garcia Fernandez, R., Vitanov, N. V., & Bergmann, K. (2003). Creation and Measurement of a Coherent Superposition of Quantum States. *Physical Review Letters*, *91*(21), 213001.

Wu, Y., Yang, L.-P., Gong, M., Zheng, Y., Deng, H., Yan, Z., Zhao, Y., Huang, K., Castellano, A. D., Munro, W. J., Nemoto, K., Zheng, D.-N., Sun, C. P., Liu, Y., Zhu, X., & Lu, L. (2018). An efficient and compact switch for quantum circuits. *Npj Quantum Information*, *4*(1), 1–8.

Qin, X., Xu, J., Wu, Y., & Liu, X. (2019). Energy-transfer editing in lanthanide-activated upconversion nanocrystals: A toolbox for emerging applications. ACS Central Science, 5(1), 29–42.

**Supplementary Note S1**

*Material use in solar cells*

As of 2022, crystalline silicon (c-Si) technology accounts for upwards of 90% of solar cell production and deployment around the world. Beside Si, other materials such as Ag and Al are used for contacts and conducting grids. According to the Photovoltaic (PV) Module Technologies Report from the National Renewable Energy Laboratory (Smith et al. 2021), dominant c-Si technologies are projected to be deployed in so-called M6 module sizes using a thickness of 160 μm. Using the density of silicon (2328 kg/m³) we can approximate material use per area:

*Material use per area (c-Si technology)* $= 160 \times 10^{-6} m \times \frac{2328 \times 10^3 g}{m^3} = 372.48 \ g/m^2$

Organic semiconductor technologies have already been demonstrated with material use of 1 g/m² (Riede et al. 2021), which is an improvement by a factor upwards of 200, even when using conservative estimations.

Another relevant unit to estimate material use is grams per Watt of power produced at peak performance (g/Wp) which is more relevant in terms of energy utilization. The advancements in c-Si have increased the energy production efficiency to approximately 3 g/Wp (Philipps & Warmuth 2022), while the current energy efficiency for commercial organic solar cells reaches between 5 and 10%, its energy output per gram is much higher than that of Si. According to reported data by the German firm Heliatek, their product HeliaSol 436-2000 with an area of (0.436 x 2 m²) can convert between 50 and 55 W of solar energy into electricity (Le Séguillon 2019).

*Material use per power output (organic semiconductor technology)* $= \frac{1 \ g}{m^2} \times \frac{0.872 \ m^2}{50 \ W} = 0.017 \ g/Wp$

In practice, many metrics need to be considered to evaluate the performance of different solar cells, including costs, degradation, lifetime and other factors. However, material use per area and material use per power output are key technical performance metrics coupled to economic usage – as defined in Magee et al. 2016 – and therefore merit particular attention.

*Power densities in batteries*

Specialized lithium-ion batteries have been shown to exhibit power densities up to 900 W/kg (Dechent et al. 2021). For higher power density applications, supercapacitors are typically used. However, supercapacitors are restricted to very low energy densities (typically below 10 Wh/kg). Quantum batteries promise to allow for high energy densities along with high power densities. The first prototype of a quantum battery, as described in Quach et al. 2022, reached a power density of about 67 kW/kg along with an energy density of about 4 Wh/kg. The example conversions are:

*Energy density* = $E_{max}$ *[eV]* × 4.45e-23 / ($A_{MM}$ × 1.661e-27) = 4.07 Wh/kg

*Power density* = $P_{max}$ *[eV/ps]* × 1.602e-7 / *approx. mass [kg]* / 1000 = 67.09 kW/kg

Compared to electrochemical batteries with maximum power densities on the order of 900 W/kg, the reported number described above represents a 74 fold enhancement.

*Energy densities in nuclear batteries*

When storing energy at the atomic level, a hard upper bound is the ionization energy, which is below 15 eV for all nongaseous elements. In other words, if more than 15 eV per atom is transferred to a solid or liquid system at the atomic level, then prompt disintegration of the receiving structure can be expected. In actuality, upper limits for energy density are often substantially lower for different atomic-level storage technologies, e.g. 1.2 eV per molecule for hydrogen fuel (see Table S1).

In contrast, nuclear excited states can store orders of magnitude higher energies. Consider, for instance, the $^{180m}$Ta nuclear isomer, which in nature exists in a metastable excited state at around 75 keV above the nuclear ground state (lifetime: ~$10^{15}$ years). This corresponds to an energy density of around 11 MWh/kg. Compare this energy density with the theoretical limit of hydrogen fuel at 33 kWh/kg. Other nuclear isotopes offer metastable excited states at the MeV range, which would entail energy densities orders of magnitude higher than that of $^{180m}$Ta (Chiara et al. 2018; Rzadkiewicz et al. 2019).

To utilize atomic nuclei as energy storage repositories, discharging mechanisms – and ideally also charging mechanisms – need to be devised. While the induced deexcitation of the metastable $^{180m}$Ta isotope, as well as of other isomers, has been demonstrated, the processes pursued thus far are energetically inefficient, e.g. excitation to unstable excited states near 1 MeV (Belic et al. 2002). The literature presented in Section 3.3 of this article describes alternative (coherent) mechanisms to potentially trigger nuclear deexcitation as well as deliberate excitation.

*Table S1. Comparison of physical energy densities across different atomic-level and nuclear-level energy storage mechanisms.*

|  | **Energy density (kWh/kg)** | **Energy Density (eV/atom)** |
|---|---:|---:|
| **Hydrogen Fuel Cell (theoretical Limit)** | 33 | 1.2411 |
| **Diesel (theoretical limit)** | 12.64 | 4.9386 |
| **Lithium-Ion (theoretical limit)** | 0.56 | N/A |
| **$^{180m}$Ta isomers** | 11104.45 | 75000 |

Conversion between the two columns of Table S1 is based on the following relationship:

*Energy density [eV/atom] = energy density [kWh/kg]* $\times \frac{1000\ Wh}{1 kWh} \times \frac{3600\ J}{1\ Wh} \times \frac{6.242E18\ eV}{1\ J} \times \frac{molar\ mass\ [kg]}{6.022E23\ atoms}$

For the case of Diesel, the composition of 86% C and 14% H was used as molar mass. Lithium-ion battery technologies were not calculated at the eV/atom level since they depend on electrochemical reactions at the cell level that involve an electrolyte besides the lithium ion species.

**Supplementary Note S2**

The quantized nature of energetic states is a hallmark of quantum systems. In other words, a quantum system can absorb and shed energy only at discrete values, which are typically characteristic of the system itself. While most quantum systems exhibit multiple energy levels, models often yield satisfactory results when focusing on only specific levels of interest. In its most simplified form, a quantum system is modeled as a two-level system (TLS) comprising an excited state and a ground state.

This setup results in one possible transition, which represents a deexcitation when the initial state is the excited state and an excitation when the initial state is the ground state. Deexcitation is considered "spontaneous" when not externally induced. Spontaneous emission follows an exponential probability distribution (Loudon 2000), whereby the mean value represents the average decay rate, also known as the half-life.

Different transition rates result when the excited state is delocalized across multiple quantum systems and if couplings via shared oscillators (fields) exist to resonant or near-resonant receiver systems. In the former case, one speaks of the collective excited state representing a superposition of excited states of the participating systems. Then, excitation and deexcitation dynamics are modified by a so-called Dicke enhancement factor (Dicke 1954). In the latter case, occupation probability can sinusoidally oscillate back and forth between the coupled systems (Rabi oscillations).

The above-described modifications to excitation and deexcitation dynamics all naturally arise from the evaluation of so-called Dicke models. While different variants of Dicke models exist, they follow the same structure, containing terms that describe the energy in one or multiple two-level-systems, the energy in one or multiple oscillator modes, and the interactions between two-level systems and oscillator modes:

$$H = \sum_j H_{tls} + H_{osc} + \sum_j H_{interaction}$$

Supporting Table S2 contains an overview of different Dicke model variants.

*Table S2: Models for single two-level-system (TLS)*

| Name | # of TLS | counter-rotating terms | Hamiltonian |
|---|---|---|---|
| Rabi | N=1 | Y | $H = \overbrace{\hbar\omega_0\sigma_z}^{tls} + \overbrace{\hbar\omega a^\dagger a}^{osc} + \overbrace{\hbar g\left(a^\dagger + a\right)\sigma_x}^{interaction}$ |
| Jaynes-Cummings | N=1 | N | $H = \hbar\omega_0\sigma_z + \hbar\omega a^\dagger a + \hbar g(\sigma_+ a + \sigma_- a^\dagger)$ |

*Table S3: Models for multiple two-level-system (TLS)*

| Name | # of TLS | counter-rotating terms | Hamiltonian |
|---|---|---|---|
| Dicke | N>1 | Y | $H = \hbar\omega_0 \sum_{j=1}^{N} \sigma_z^j + \hbar\omega a^\dagger a + \hbar g \sum_{j} \sigma_x^j(a + a^\dagger)$ |
| Tavis-Cummings | N>1 | N | $H = \hbar\omega_0 \sum_{j=1}^{N} \sigma_z^j + \hbar\omega a^\dagger a + \hbar g \sum_{j} (\sigma_+^j a + \sigma_-^j a^\dagger)$ |

**Supplementary Note S3**

In Lloyd and Mohseni 2010, a Dicke model is given in Eq. (1):

$$H = \hbar\omega a^\dagger a - \hbar\frac{\omega}{2} \sum_{j=1}^{N} \sigma_z^j + \hbar\gamma \sum_{j=1}^{N} \sigma_x^j(a + a^\dagger)$$

This model is then specified to estimate the rate of excitation transfer between collective excitation spread out across *N* acceptor sites (A sites) to *M* acceptor sites (B sites):

$$H = -\hbar\frac{\omega_A}{2} \sum_{j=1}^{N} \sigma_z^j - \hbar\frac{\omega_B}{2} \sum_{k=1}^{M} \sigma_z^k + \hbar\gamma \sum_{j=1,k=1}^{N,M} (\sigma_+^j \sigma_-^k + \sigma_-^j \sigma_+^k)$$

As an example for a physical system, a network of chromophore molecules is considered, as in the case in photosynthesis. The excitation transfer rate then increases as a function of the number of participating molecules. As stated by the authors:

> *"A closely packed group of N molecules under certain symmetry can collectively accept or donate an excitation with a rate that is almost N times faster than each individual molecule."*

In other words, the transition rate is then not just determined by the coupling *g* (here given as *γ*) but also by the number of participating systems *N*. The authors refer to this enhancement effect in the excitation transfer rate as *supertransfer*.

In Abasto et al. 2012, an equivalent Hamiltonian is given, with slightly different notation:

$$H = -\frac{\epsilon_A}{2} \sum_{i=1}^{n_A} \sigma_z^i - \frac{\epsilon_B}{2} \sum_{j=1}^{n_B} \sigma_z^j + \gamma \sum_{i=1,j=1}^{n_A,n_B} \sigma_+^i \sigma_-^j + \sigma_-^i \sigma_+^j$$

Here, the authors elaborate similarly to the above:

> *"Under symmetrized interactions of a group of n molecules, the excitation becomes highly delocalized, leading to a large (effective) dipole moment associated with the entire group. The resulting enhanced oscillator strength can lead to supertransfer when similar molecular assemblies, with comparable effective dipole moments, exist that can play the role of acceptors.*

> *Under such conditions, the rate of exciton dynamics should be calculated from these effective large dipole–dipole interactions to describe the coherent donation and acceptance among such molecular aggregates, with up to $n^2$ enhancement over the single molecule to single molecule transfer rate, even in the far field."*

In an additional twist, the superposition stage in nonradiative excitation transfer can not only be utilized for transferring energy but also for converting it. Specifically, if an incoming photon carries energy that exceeds the bandgap of the absorbing material – e.g. in the ultraviolet regime – then this excess energy is lost. However, if the large exciton energy is transferred to not one but two sites for charge generation, then the available photon energy can be more completely utilized. This quantum-based energy conversion technique is known as *exciton fission* (Rao & Friend 2017; Einzinger et al. 2019).

**Supplementary Note S4**

In Quach et al. 2022, a Dicke model is given in Eq. (3):

$$H = \frac{\hbar(\omega - \omega_L)}{2} \sum_{j=1}^{N} \sigma_j^z + \hbar(\omega - \omega_L)a^\dagger a + g \sum_{j=1}^{N}(a^\dagger \sigma_j^- + a\sigma_j^+) + i\hbar\eta(t)(a^\dagger - a)$$

where $\Delta = \omega - \omega_L$ describes detuning and where $\eta(t)$ is the Gaussian pulse envelope of the laser:

$$\eta(t) = \frac{\eta_0}{\sigma\sqrt{2\pi}} e^{-\frac{1}{2}(\frac{t-t_0}{\sigma})^2}$$

Here, organic Lumogen-F orange molecules are used as two-level systems. N denotes how many of such systems are coupled. When $\Delta = \omega - \omega_L = 0$, then the laser is on resonance.

In their supplementary materials, the authors show that excitation and deexcitation times in such systems can grow as a function of $g\sqrt{N}$.

**Supplementary Note S5**

In Haber et al. 2017, a Dicke model is given in Eq. (1):

$$\begin{aligned} H = {}& \Delta_1 a_1^\dagger a_1 + \Delta_2 a_2^\dagger a_2 + J(a_1^\dagger a_2 + a_2^\dagger a_1) \\ & - \Delta(|E_1\rangle\langle E_1| + |E_2\rangle\langle E_2|) \\ & + g_1\sqrt{N_1}(a_1|E_1\rangle\langle G| + a_1^\dagger|G\rangle\langle E_1|) \\ & + g_2\sqrt{N_2}(a_2|E_2\rangle\langle G| + a_2^\dagger|G\rangle\langle E_2|) \end{aligned}$$

where *G* and *E* denote nuclear ground and excited states respectively.

In the modeled systems, two cavities exist (described by operators $a_1$ and $a_2$), each of which couples to an ensemble of $^{57}$Fe nuclei (with excited states $E_1$ and $E_2$) whereby $g_1$ and $g_2$ are the respective coupling strengths between each cavity and its ensemble. Note that de facto $g_1$ and $g_2$ scale with the square root of the number of nuclei in each ensemble ($N_1$ and $N_2$).

$\Delta$ is the detuning between X-rays and the nuclear resonance energy.

From the above, a coupling term that describes the interaction between the two ensembles of nuclei represented by $E_1$ and $E_2$ can be derived. It is given in Haber et al. 2017 as:

$$g_{12} = -\frac{g_1 g_2 N}{2}\left(\frac{\Delta_+}{\kappa_+^2 + \Delta_+^2} - \frac{\kappa_-}{\Delta_-^2 + \Delta_-^2}\right)$$

where $\kappa$ denotes the X-ray driving strength.

**Supplementary Note S6**

In Hagelstein & Chaudhary 2015b and Metzler et al. 2022, a Dicke model is given as Eq. (1):

$$H = \overbrace{\sum_j M_j c^2}^{nuclei} + \overbrace{\hbar\omega a^\dagger a}^{osc} + \overbrace{\sum_j (-\mu_j \cdot B)_+ + (-\mu_j \cdot B)_-}^{magn.\,interaction}$$

with $M_j$ containing two nuclear states:

$$M_j = \begin{bmatrix} M_1 c^2 & 0 \\ 0 & M_2 c^2 \end{bmatrix}$$

In Metzler et al. 2022, excitation transfer rate estimates are given for the case of donors comprising an ensemble of $D_2$ nuclei and acceptors comprising an ensemble of $^4$He nuclei.

The excitation transfer rate is then given as (Eq. S17):

$$\Gamma = \frac{\frac{[ge^{-G}\sqrt{\frac{vol_{nuc}}{vol_{mol}}}]g}{\Delta E}\sqrt{N_{D_2}}\sqrt{N_{He}}}{\hbar}$$

with a volumetric factor

$$\frac{vol_{nuc}}{vol_{mol}} = \frac{\frac{4}{3}\pi r_{nuc}^3}{2\pi^2 R_0 \Delta R^2} = 6.26 \times 10^{-12}$$

that accounts for volumetric change in the course of the transition and the Gamow factor

$$e^{-G}$$

that denotes the hindrance to the transition represented by Coulombic repulsion.

Moreover, $\sqrt{N_{D_2}}\sqrt{N_{He}}$ represents the Dicke enhancement factor, which is a function of the number of participating nuclei.

Finally, $g = -\mu \cdot B$ where $\mu = 3.15 \times 10^{-8}\ eV\ T^{-1}$ is the nuclear magneton and $B$ is the internal magnetic field affecting nuclei in a lattice. For $B$ on the order of 3 T, an estimate for $g$ on the order of $10^{-7}$ eV results.

This leads to an estimated enhancement of the $|D_2> \rightarrow |^4He>$ transition by about 30 orders of magnitude (Eq. S18 in Metzler et al. 2022), compared with the incoherent transition rate of $10^{-64}$/s:

$$\Gamma = \frac{10^{-7}eV\,10^{-33}10^{-6}10^{-7}eV}{24 \times 10^6 eV} 1 \sqrt{10^{12}}\sqrt{10^6}\frac{1}{\hbar} = 10^{-34}s^{-1}$$

Analogous to the principle of exciton fission that is utilized in quantum solar engineering (see Supplementary Note S3), energy conversion via superposition states is proposed to apply to quantum nuclear engineering as well. For instance, Hagelstein & Chaudhary 2015a suggest that a single high-energy nuclear state transition such as a fusion reaction may resonantly trigger multiple lower-energy nuclear state transitions such as excitation in multiple coupled acceptor nuclei.

**References Supplementary Notes:**

Abasto, D. F., Mohseni, M., Lloyd, S., & Zanardi, P. (2012). Exciton diffusion length in complex quantum systems: The effects of disorder and environmental fluctuations on symmetry-enhanced supertransfer. *Philosophical Transactions of the Royal Society A: Mathematical, Physical and Engineering Sciences*, *370*(1972), 3750–3770.

Belic, D., Arlandini, C., Besserer, J., Boer, J., Carroll, J., Enders, J., Hartmann, T., Kaeppeler, F., Kaiser, H., Kneissl, U., Kolbe, E., Langanke, K., Loewe, M., Maier, H., Maser, H., Mohr, P., Von Neumann-Cosel, P., Nord, A., Pitz, H., & Zilges, A. (2002). Photo-induced depopulation of the$^{180}$ Ta$^{m}$ isomer via low-lying intermediate states: Structure and astrophysical implications. *Phys. Rev. C*, *65*.

Chiara, C. J., Carroll, J. J., Carpenter, M. P., Greene, J. P., Hartley, D. J., Janssens, R. V. F., Lane, G. J., Marsh, J. C., Matters, D. A., Polasik, M., Rzadkiewicz, J., Seweryniak, D., Zhu, S., Bottoni, S., Hayes, A. B., & Karamian, S. A. (2018). Isomer depletion as experimental evidence of nuclear excitation by electron capture. *Nature*, *554*(7691), Article 7691.

Dechent, P., Epp, A., Jost, D., Preger, Y., Attia, P. M., Li, W., & Sauer, D. U. (2021). ENPOLITE: Comparing lithium-ion cells across energy, power, lifetime, and temperature. *ACS Energy Letters*, *6*(6), 2351–2355.

Dicke, R. H. (1954). Theory of superradiance. Physical Review, 93(1), 99–110.

Einzinger, M., Wu, T., Kompalla, J. F., Smith, H. L., Perkinson, C. F., Nienhaus, L., Wieghold, S., Congreve, D. N., Kahn, A., Bawendi, M. G., & Baldo, M. A. (2019). Sensitization of silicon by singlet exciton fission in tetracene. *Nature*, *571*(7763), Article 7763.

Haber, J., Kong, X., Strohm, C., Willing, S., Gollwitzer, J., Bocklage, L., Rüffer, R., Pálffy, A., & Röhlsberger, R. (2017). Rabi oscillations of X-ray radiation between two nuclear ensembles. *Nature Photonics*, *11*(11), Article 11.

Hagelstein, P. L., & Chaudhary, I. U. (2015a). Anomalies in fracture experiments, and energy exchange between vibrations and nuclei. *Meccanica*, *50*(5), 1189–1203.

Hagelstein, P. L., & Chaudhary, I. U. (2015b). Phonon models for anomalies in condensed matter nuclear science. *Current Science* (00113891), 108(4), 507–513.

Le Séguillon, T. (2019). Heliatek: Les films photovoltaïques, de la paillasse à l'usine. *Le Journal de l'ecole de Paris Du Management*, *138*(4), 30–36.

Lloyd, S., & Mohseni, M. (2010). Symmetry-enhanced supertransfer of delocalized quantum states. *New Journal of Physics*, *12*(7), 075020.

Loudon, R. (2000). *The quantum theory of light*. OUP Oxford.

Magee, C. L., Basnet, S., Funk, J. L., & Benson, C. L. (2016). Quantitative empirical trends in technical performance. *Technological Forecasting and Social Change*, *104*, 237–246.

Metzler, F., Hunt, C., & Galvanetto, N. (2022). Known mechanisms that increase nuclear fusion rates in the solid-state (arXiv:2208.07245). *arXiv*. https://doi.org/10.48550/arXiv.2208.07245

Philipps, S., & Warmuth, W. (2022). *Photovoltaics Report Fraunhofer ISE*. Fraunhofer Institute for Solar Energy Systems ISE, Freiburg im Breisgau, Germany.

Quach, J. Q., McGhee, K. E., Ganzer, L., Rouse, D. M., Lovett, B. W., Gauger, E. M., Keeling, J., Cerullo, G., Lidzey, D. G., & Virgili, T. (2022). Superabsorption in an organic microcavity: Toward a quantum battery. *Science Advances*, *8*(2), eabk3160.

Rao, A., & Friend, R. H. (2017). Harnessing singlet exciton fission to break the Shockley–Queisser limit. *Nature Reviews Materials*, *2*(11), Article 11.

Riede, M., Spoltore, D., & Leo, K. (2021). Organic Solar Cells—The Path to Commercial Success. *Advanced Energy Materials*, *11*(1), 2002653.

Rzadkiewicz, J., Polasik, M., Słabkowska, K., Syrocki, Ł., Węder, E., Carroll, J. J., & Chiara, C. J. (2019). Beam-based scenario for $^{242m}\mathrm{Am}$ isomer depletion via nuclear excitation by electron capture. *Physical Review C*, *99*(4), 044309.

Smith, B. L., Woodhouse, M., Horowitz, K. A. W., Silverman, T. J., Zuboy, J., & Margolis, R. M. (2021). *Photovoltaic (PV) Module Technologies: 2020 Benchmark Costs and Technology Evolution Framework Results* (NREL/TP-7A40-78173). National Renewable Energy Lab. (NREL), Golden, CO (United States).